\documentclass[journal]{IEEEtran}
\usepackage[utf8]{inputenc}

\usepackage[dvipsnames, table]{xcolor}

\usepackage[hidelinks]{hyperref}
\usepackage{cite}
\usepackage{microtype}
\usepackage[nolist]{acronym}
\usepackage{dirtytalk}
\usepackage{pifont}
\usepackage{gensymb}
\usepackage{orcidlink}
\usepackage{booktabs, dcolumn}
\usepackage{multirow}

\usepackage{bm}
\usepackage{mathtools}
\usepackage{physics, amsmath,amssymb,amsfonts}

\usepackage{caption}
\usepackage{subcaption}
\usepackage[export]{adjustbox}
\usepackage{graphicx}
\usepackage{pgfplots} 
\usepackage{placeins}
\DeclareUnicodeCharacter{2212}{−}
\usepgfplotslibrary{groupplots,dateplot}
\usetikzlibrary{quotes,arrows.meta, calc, patterns.meta, shapes.multipart, shapes.arrows, shapes.misc, spy, decorations.pathreplacing, arrows.meta, backgrounds, shapes.geometric, intersections, matrix, fadings, patterns}
\pgfplotsset{compat=1.15}
\usepackage{tikzpagenodes} %

\def\BibTeX{{\rm B\kern-.05em{\sc i\kern-.025em b}\kern-.08em
    T\kern-.1667em\lower.7ex\hbox{E}\kern-.125emX}}

\begin{acronym}
\acro{DFT}{discrete Fourier transform}
\acro{MVDR}{minimum variance distortionless response}
\acro{PDF}{probability density function}
\acro{MMSE}{minimum mean square error}
\acro{ML}{maximum likelihood}
\acro{SNR}{signal-to-noise ratio}
\acro{MAP}{maximum a posteriori}
\acro{ASR}{automatic speech recognition}
\acro{POLQA}{perceptual objective listening quality analysis}
\acro{MOS}{mean opinion score}
\acro{PESQ}{perceptual evaluation of speech quality}
\acro{EM}{expectation maximization}
\acro{DNN}{deep neural network}
\acro{LSTM}{long short-term memory}
\acro{FF}{feed-forward}
\acro{cIRM}{complex ideal ratio mask}
\acro{IRM}{ideal ratio mask}
\acro{RIR}{room impulse response}
\acro{ATF}{acoustic transfer function}
\acro{RTF}{relative transfer function}
\acro{CRNN}{convolutional recurrent neural network}
\acro{CNN}{convolutional neural network}
\acro{STFT}{short-time Fourier transform}
\acro{CQS}{continuous quality scale}
\acro{RIR}{room impulse response}
\acro{DoA}{direction of arrival}
\acrodefplural{DoA}{directions of arrival}
\acro{DRR}{direct-to-reverberation ratio}
\acro{PIT}{permutation invariant training}
\acro{SI-SDR}{scale-invariant source-to-distortion ratio}
\acro{TdoA}{time differences of arrival}
\acrodefplural{TdoA}{time differences of arrival}
\acro{IPD}{inter-channel phase difference}
\acrodefplural{IPD}{inter-channel phase differences}
\end{acronym}

\newcommand{\numMics}{\ensuremath{C}}

\newcommand{\numSpeaker}{\ensuremath{P}}
\newcommand{\spidx}{\ensuremath{p}}
\newcommand{\micidx}{\ensuremath{\ell}}
\newcommand{\sampleidx}{\ensuremath{t}}
\newcommand{\freqbinidx}{\ensuremath{k}}
\newcommand{\timeframeidx}{\ensuremath{i}}

\begin{document}
\title{Multi-channel Speech Separation Using Spatially Selective Deep Non-linear Filters}

\author{Kristina~Tesch\,{\orcidlink{0000-0002-6458-8128}},~\IEEEmembership{Student Member,~IEEE}, and
        Timo~Gerkmann\,{\orcidlink{0000-0002-8678-4699}},~\IEEEmembership{Senior Member,~IEEE}%
\thanks{The authors are with the Signal Processing Group, Department of Informatics, Universität Hamburg, 22527 Hamburg, Germany (e-mail: kristina.tesch@uni-hamburg.de; timo.gerkmann@uni-hamburg.de).\\%
This work was funded by the Deutsche Forschungsgemeinschaft (DFG, German Research Foundation) — project number 508337379.}%
}

\maketitle

\begin{tikzpicture}[remember picture,overlay]
  \node [draw=black, fill=white, text width=\textwidth, inner sep=2pt, yshift=-0.9cm, execute at begin node=\setlength{\baselineskip}{2ex}] at (current page text area.south){\footnotesize{Accepted paper. \copyright 2023 IEEE. Personal use of this material is permitted. Permission from IEEE must be obtained for all other uses, in any current or future media, including reprinting/republishing this material for advertising or promotional purposes, creating new collective works, for resale or redistribution to servers or lists, or reuse of any copyrighted component of this work in other works.}};
\end{tikzpicture}

\begin{abstract}
In a multi-channel separation task with multiple speakers, we aim to recover all individual speech signals from the mixture.
In contrast to single-channel approaches, which rely on the different spectro-temporal characteristics of the speech signals, multi-channel approaches should additionally utilize the different spatial locations of the sources for a more powerful separation especially when the number of sources increases. To enhance the spatial processing in a multi-channel source separation scenario, in this work, we propose a \ac{DNN} based spatially selective filter (SSF) that can be spatially steered to extract the speaker of interest by initializing a recurrent neural network layer with the target direction. We compare the proposed SSF with a common end-to-end direct separation (DS) approach trained using utterance-wise \ac{PIT}, which only implicitly learns to perform spatial filtering. We show that the SSF has a clear advantage over a DS approach with the same underlying network architecture when there are more than two speakers in the mixture, which can be attributed to a better use of the spatial information. Furthermore, we find that the SSF generalizes much better to additional noise sources that were not seen during training and to scenarios with speakers positioned at a similar angle.
\end{abstract}


\section{Introduction}
Speech separation algorithms target the so-called cocktail party problem, where several (two or more) human speakers are speaking at the same time. The goal is to recover the original speech signals from a mixture recording that may also contain additional background noise and reverberation. This task is particularly challenging because all target speech signals have similar tempo-spectral characteristics. But nevertheless, normal-hearing people are very good at focusing their attention on a single target speaker, so that they can even enjoy a conversation at a cocktail party. This ability is mainly due to the fact that humans have two ears, which enables them to perceive and process spatial information. Similarly, also speech processing algorithms can leverage spatial information by using multiple microphones to record the mixture signals. 

The most traditional form of spatial processing is to employ a linear spatial filter, a so-called beamformer, which is designed to enhance the signal arriving from a target direction by suppressing signal components that arrive from a direction other than the target direction. Two prominent examples are the Delay-and-Sum beamformer and the \ac{MVDR} beamformer \cite{vary2006digital}. Both of these employ a linear processing model: first, the individual microphone signals are filtered, and then added. The underlying idea for the Delay-and-Sum beamformer is to compensate for the relative \acp{TdoA} of the signal at the microphone channels in the filtering step. Therefore, accurate \ac{TdoA} or related \ac{DoA} estimates are required. The \ac{MVDR} beamformer additionally takes the second-order statistics of the interfering signal into account so that it can form a superdirective beamformer or steer nulls in the direction of interfering point sources. However, the number of point-sources that can be eliminated is bounded by the number of microphone channels minus one \cite[Sec. 6.3]{trees2004OptimumArrayProcessing}. Consequently, the \ac{MVDR} beamformer deteriorates in a reverberant setting as reflections of the interfering sources arrive from all directions. As the performance of these linear spatial filters is limited, a single-channel post-filter is commonly applied to the output of the linear spatial filter. 

Linear spatial filters have been, and still are, a popular choice for source separation problems because of their ability to focus on a single target source based on its spatial characteristics \cite{madhu2011, markovich2009eigenspacebeam, souden2013framew, 2017gannot, 2017taseskadoa, drude2017tightintegration, yoshioka2018farfieldrec, wang2021multiframesep, sivasankaran2021, liu2022new, subramanian2020, chen2018beamintegration, aroudi2021dbnet}. The challenging part here is accurate parameter estimation: a data-dependent implementation of the \ac{MVDR} requires the localization of speakers or a direct estimation of the \ac{RTF} as well as an estimation of the interfering signal's covariance matrix. While older works employed statistical modeling, e.g., \cite{souden2013framew, 2017taseskadoa}, recent ones rely on neural networks for this purpose. In the so-called masked-based beamforming approach, a neural network is used to estimate time-frequency masks for each speaker and use these to obtain the target speech and interfering noise signal's covariance matrix, e.g., \cite{heymann2015blstm, sivasankaran2021, wang2021multiframesep, subramanian2020}. Other researchers suggest to sample the space with a fixed beamformer and use a neural network for beam selection and post-filtering \cite{aroudi2021dbnet, liu2022new, chen2018beamintegration}. 

However, with the rise of the neural network era, there is also an increasing number of multi-channel speech separation approaches that do not perform explicit spatial filtering. Instead, the neural networks are presented with multi-channel inputs and/or directional features and are trained to estimate the speech sources directly from the mixture, e.g., \cite{wang2018mcdeepclustering, gu2020spatialfeature, luo2020fastacnet, quan2022nbconformer}. Throughout this work, we will refer to these end-to-end regression-based systems as direct separation (DS) approaches. Typically, these systems output as many speech signals as there are speech sources in the mixture, which gives rise to a permutation problem. For this reason, most DS approaches are trained with an utterance-wise \acf{PIT} loss. Unlike in the case of using a beamformer for spatial filtering, the spatial processing takes place only implicitly in the DS networks. It is clear, however, that maximum separation performance can only be reached if such a network learns to perform powerful spatial processing directly from training examples.

While the traditional spatial filters are constraint by a linear processing model, the nature of neural networks enables non-linear spatial processing and, furthermore, an integration of the spatial and tempo-spectral processing steps, which are separated in a traditional beamformer plus tempo-spectral post-filter setup. In a previous analysis based on statistical \ac{MMSE} estimators \cite{tesch2021nonlinearspatialfilteringtasl}, we have shown that this indeed leads to more powerful spatial processing in non-Gaussian interferences, which is arguably always the case in speech separation. As a consequence, the upper bound for the number of sources that can be canceled by a linear spatial filter does not hold anymore if the filter is non-linear and jointly performs spatial and tempo-spectral processing. In further experiments with \ac{DNN}-based joint non-linear spatial and tempo-spectral filters, we have confirmed that neural networks can implement spatial filters that drastically outperform an oracle \ac{MVDR} plus additional \ac{DNN}-driven post-filter \cite{tesch2022tasl}. Accordingly, DNN-based DS approaches can offer a potentially better spatial filtering than a traditional linear spatial filter. On the other hand, these networks have to learn the spatial processing implicitly from data. How well this is accomplished can only be determined indirectly. Since the DS approaches are less modular than the spatial filtering approach that separates parameter estimation, e.g., \ac{DoA} estimation, and spatial filtering, they are also less flexible with respect to, for example, a variable number of sources. In this paper, we investigate the separation performance of DNN-based non-linear joint spatial and tempo-spectral filters, which have been trained according to two different strategies: (1) using \ac{PIT}, which means that the spatial filtering must be learned implicitly from the provided examples and (2) with an explicit focus on the spatial filtering by steering a filter towards a target speaker with a given DoA. A filter obtained by using the second strategy is referred to as a spatially selective filter (SSF) in this work. 

Many researchers have proposed to enhance multi-channel speech separation with so-called directional features \cite{wang2018mcdeepclustering, wang2019spatialspectralfeatures} or use location-information to guide speaker extraction tasks \cite{gu2020spatialfeature, gu2019neuralspatialfilter, 2018chenlocationguided}. For example, Gu et al. and \cite{gu2019neuralspatialfilter} and Chen et al. \cite{2018chenlocationguided} proposed an angle feature indicating which time-frequency bins are dominated by a signal from a particular \ac{DoA}. Other common features are related to the \acp{IPD} of the microphone pairs, cross-correlation features or features computed with fixed beamformers, e.g., a Delay-and-Sum beamformer steered to a set of candidate locations \cite{2018chenlocationguided, liu2022new}. In contrast, in our proposed approach, we do not rely on hand-crafted features, but use a neural network to learn the spatial processing from raw multi-channel data. 

In our recent ICASSP 2023 paper \cite{tesch2023icassp}, we have introduced a conditioning mechanism to flexibly steer a DNN-based non-linear spatial filter in a desired target direction. Given the noisy mixture and the target look-direction of the filter, the SSF then extracts the speech signal corresponding to the speaker located in that direction, similar to traditional linear spatial filters. This ability to flexibly steer the filter in a desired direction is a major improvement over a filter with a fixed look-direction \cite{tesch2022tasl, tesch2022interspeech}, or with fixed spatial target regions \cite{markovic2022nsfmeta,wechsler2023spatialregions}. The conditioning mechanism we proposed in \cite{tesch2023icassp} does not need a steering vector like a classic linear beamformer or the related work by Jenrungrot et al. \cite{jenrungrot2020cos} but is conditioned on the one-hot encoded angle. This avoids an implicit far-field assumption and leads to better performance, as we showed in \cite{tesch2023icassp}. Similarly, Kindt et al. \cite{kindt2022closespeakers} have shown that a learned encoding based on a one-hot encoded angle used as a feature to improve separation of closely spaced speakers is more valuable than a hand-crafted feature based on expected phase differences.

In this paper, we extend our previous work and investigate the use of SSFs for speech separation. We aim to understand if the explicit spatial filtering in SSFs is advantageous over the implicit spatial filtering learned by the widely adopted DS approach in terms of overall performance, but also in terms of generalization ability to conditions unseen during training. Furthermore, we investigate the robustness of the SSF to errors in the \ac{DoA} input as well as pertubations in the microphone array geometry.  

The rest of the paper is structured as follows: In the next section, we give a formal problem description and introduce the notation. In Section \ref{ssec:netarc}, we describe two neural network architectures for joint spatial and tempo-spectral non-linear filtering, which we use to compare a DS and SSF approach using the same underlying network architecture. In addition, we explain the steering mechanism of the SSFs in Section \ref{ssec:conditioning}. Section \ref{sec:data} describes the dataset generation, and in Section \ref{sec:sepperf}, we compare the speech separation performance of the two approaches. Detailed investigations on the robustness and generalization ability are presented in Section \ref{sec:robustness} and Section \ref{sec:generalization}.

\section{Problem Definition}
In this work, we consider a multi-channel reverberant speech separation scenario. The goal is to recover the speech signals uttered by $\numSpeaker$ concurrently speaking persons in a reverberant room. The mixture signal is recorded by an omni-directional microphone array with $\numMics$ channels. We denote the dry speech signal of the $\spidx$'s speaker by $s_\spidx(\sampleidx)$ with time-index $\sampleidx$. The recording of $s_\spidx(\sampleidx)$ at the $\micidx$'s microphone includes not only a time-shift due to the propagation delay but also reflections on the walls and is denoted by $x_\spidx^\micidx (\sampleidx)$. Given the \ac{RIR} $h_\spidx^\micidx(\sampleidx)$ describing the propagation path of the signal uttered by the $\spidx$ speaker to the $\micidx$'s microphone, the dry and recorded signal are related via a convolution operation, i.e., 
\begin{equation}
    x_\spidx^\micidx (\sampleidx) = s_\spidx(\sampleidx) * h_\spidx^\micidx(\sampleidx).
\end{equation}
Using the \ac{STFT}, we transform the time-domain signals $x_\spidx^\micidx (\sampleidx)$ into their complex-valued frequency-domain representations $X_\spidx^\micidx (\freqbinidx, \timeframeidx)$ with frequency-bin index $\freqbinidx$ and time-frame index $\timeframeidx$. The letters $F$ and $T$ denote the total number of frequency bins and time frames respectively. Following the additive signal model, the observed mixture signal is then given by 
\begin{equation}
    Y^\micidx(k,i) = \sum_{p=1}^P X_\spidx^\micidx (\freqbinidx, \timeframeidx) + V^\micidx(k,i), 
\end{equation}
where $V^\micidx(k,i)$ denotes the sensor and environmental noise possibly recorded at the $\micidx$'s microphone in addition to the speech signals. Given the mixture signal, the task now is to recover the original speech signals $S_p(k,i)$ or, equivalently $s_p(t)$, except the propagation delay caused by the length of the direct path between source and microphone array. Accordingly, we use the direct-path dry speech signals as training target and to compute metrics that require a reference signal.

\section{Spatially Selective Non-Linear Filter (SSF)}
In this work, we investigate the use of a spatially selective deep non-linear filter (SSF) for multi-channel speech separation. Our proposed method is in line with the common approach to separate the localization task from the actual speaker extraction, which is then performed in a second step using a spatial filter steered to the speaker locations. In this section, we describe two network architectures for joint spatial and tempo-spectral non-linear filtering (JNF \cite{tesch2022interspeech, tesch2022tasl} and McNet \cite{quan2022nbconformer}) and explain the proposed mechanism for flexibly steering the filter in the desired target direction.

\subsection{Network architectures for joint spatial and tempo-spectral non-linear filtering}\label{ssec:netarc}

\begin{figure}
    \centering
    \begin{subfigure}[b]{\columnwidth}
         \centering
         \includegraphics{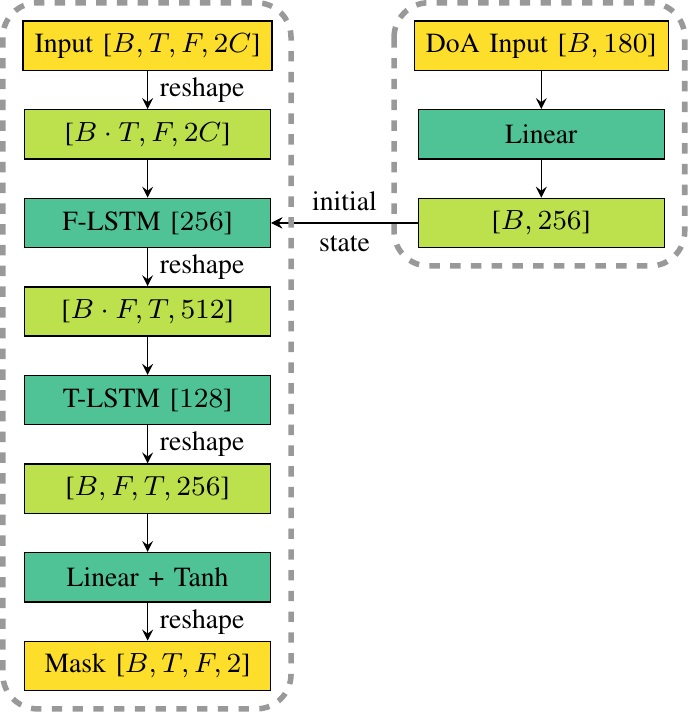}
         \caption{JNF \cite{tesch2022interspeech, tesch2022tasl} architecture (left) with steering mechanism (right)}
         \label{fig:jnfarc}
     \end{subfigure}
    \begin{subfigure}[b]{\columnwidth}
         \centering
         \vspace{1em}
         \includegraphics{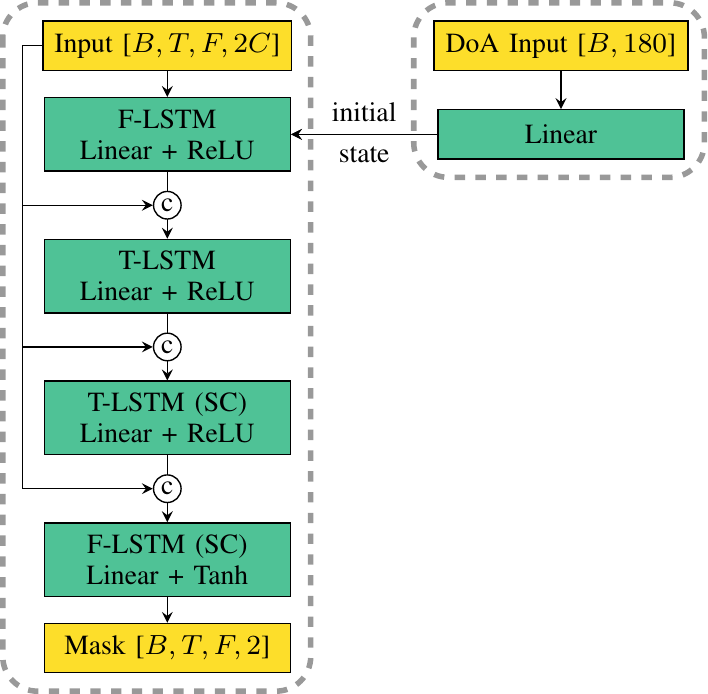}
         \caption{McNet \cite{yang2023mcfuse} architecture (left) with steering mechanism (right)}
         \label{fig:mcnetarc}
     \end{subfigure}
    \caption{Schematic view of a spatially selective filter (SSF) based on the JNF (top) and McNet (bottom) network architecture. The proposed conditioning on the target DoA is depicted on the right side.}
    \label{fig:netarc}
\end{figure}

In our prior work \cite{tesch2022interspeech, tesch2022tasl}, we have proposed the FT-JNF architecture, which we refer to as JNF in this work. The network architecture is depicted on the left side of Figure \ref{fig:jnfarc}. The JNF network expects a three-dimensional frequency-domain input. The yellow box on the top left of Figure \ref{fig:jnfarc} visualizes the input of the filter including the batch dimension denoted by $B$. The last dimension expects the real and imaginary part of all $C$ channels stacked into a vector of length $2C$. The multi-channel input provides three sources of information, which should be exploited by the network: spatial, spectral and temporal information. For this, we have previously proposed the depicted architecture with two LSTM layers at its core. The F-LSTM has been designed to extract features related to spatial and spectral information as well as their relationship, while excluding temporal correlations. This is achieved by a rearrangement of the data, which moves the time dimension into the batch dimension such that all time-frames are processed independently with network weights shared across all time-frames. The reshaping operations on the data are represented by the light green boxes in Figure \ref{fig:jnfarc}. The correlations along the time axis are then processed by the T-LSTM layer, which performs independent processing of all frequency-bins. The design has been inspired by the work of Li and Horaud \cite{li2019narrowband}, who propose a network that stacks two T-LSTM layers. However, the replacement of the first T-LSTM by the F-LSTM significantly enhances the spatial selectivity of the resulting filter such that the JNF outperformed other state-of-the-art methods on a speaker extraction task in \cite{tesch2022tasl}. 

The JNF outputs a compressed complex-valued mask in the range $[-1, 1]$, which is expanded into an uncompressed mask following the description in \cite{williamson2016cirm} with the steepness parameter set to one to obtain the single-channel time-frequency mask $\mathcal{M}_p(k,i)$ for the $p$th speaker located in the direction the filter was steered toward. The corresponding estimate $\hat{S}_p$ is given by multiplication of the mask with the reference channel of the noisy observation, i.e., 
\begin{equation}\label{eq:maskappl}
    \hat{S}_p(k,i) = Y^0(k,i) \cdot \mathcal{M}_p(k,i).
\end{equation}

The successful combination of different sources of information in the JNF architecture \cite{tesch2022tasl}, has inspired Yang et al. \cite{yang2023mcfuse} to improve it further by appending two more LSTM layers that are focused on the single-channel (SC) spectral correlations in time and frequency dimension. A schematic view of the resulting network architecture, named McNet, is shown in Figure \ref{fig:mcnetarc}. Besides two additional LSTM layers, the authors have introduced skip connections and add additional feed-forward layers after every LSTM layer. The first skip connection concatenates the noisy multi-channel signal to the input of the T-LSTM and the second and third skip connection concatenates the noisy magnitude of the reference channel to the input of the two single-channel LSTM layers. Please refer to \cite{yang2023mcfuse} for a more detailed illustration, which also includes the reshaping steps for McNet. For all experiments, we use the default configuration of McNet. Since the steering mechanism, proposed in the next section, targets the first F-LSTM layer, which is the same in both networks, we can steer both DNN-based filters in the same way and perform experiments with a spatially selective filter based on the JNF architecture (JNF-SSF) and based on the McNet architecture (McNet-SSF). 

\subsection{Proposed steering of the non-linear spatial filter (conditioning on target direction)}\label{ssec:conditioning}

In addition to the multichannel signal input, the proposed SSF requires the steering direction as a second input, as shown on the right side of Figure \ref{fig:jnfarc} and Figure \ref{fig:mcnetarc}. The direction information is presented to the network as a one-hot encoded vector, whose dimension depends on the chosen angular resolution. Figure \ref{fig:netarc} illustrates a $2^\circ$ angle resolution, which leads to $180$ possible input vectors. The one-hot vector is then fed into a linear layer, which provides an encoding of the direction information that matches with the number of units in the F-LSTM layer, which we set to $256$ in the JNF architecture. The encoded \ac{DoA} information is then used to initialize the forward and backward initial states of the bi-directional F-LSTM layer. A similar conditioning mechanism has also been used by Vinyals et al. \cite{vinyals2015initstate} to initialize a network for image caption generation with information about the image.

In contrast to our previous paper \cite{tesch2023icassp}, we only initialize the first F-LSTM layer with the direction information and omit this step for the second T-LSTM layer. Preliminary experiments have shown that conditioning the first LSTM layer leads to much better performance than conditioning the second LSTM. Furthermore, we observed that conditioning both layers does not provide a benefit but slightly increases the computational demands. These findings are in line with our previous observations in \cite{tesch2022tasl} that the spatial selectivity is mainly controlled by the F-LSTM layer.

In \cite{tesch2023icassp}, we compared the proposed conditioning mechanism based on a one-hot angle encoding to the method suggested by Jenrungrot et al. \cite{jenrungrot2020cos}: using knowledge of the microphone array geometry and based on a far-field assumption, the individual input channels are shifted such that signals arriving from the target \ac{DoA} are time-aligned. These time-aligned signals are then used as input signal of the network. We found that the target speaker is reliably extracted with this competing method, but the proposed conditioning on the encoded \ac{DoA} angle consistently performs better and does not rely on a far-field assumption.

\section{Dataset Generation}
\label{sec:data}
\begin{figure}
    \centering
    \includegraphics{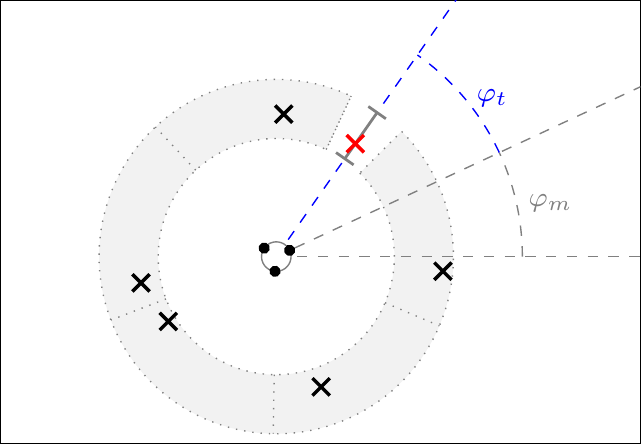}
    \caption{Illustration of the dataset generation. The target source is marked with a red cross and its DoA angle $\varphi_t$ is computed relative to the microphone orientation in the room given by $\varphi_m$. Interfering sources are placed in the gray area.}
    \label{fig:simudata}
\end{figure}

\begin{table}
    \caption{The room characteristics are sampled uniformly from the displayed ranges.}
    \label{table:simuparams}
    \centering
    \begin{tabular}{cccc}\toprule
         Width & Length & Height & $\text{T}_{60}$ \\\midrule
          $2.5-5$ m &  $3-9$ m  &  $2.2-3.5$ m   & $0.2 - 0.5$ s \\\bottomrule
    \end{tabular}
\end{table}

Using \texttt{pyroomacoustics} \cite{scheibler2018pyroomacoustics}, we generate a simulated dataset for training and evaluation based on the image-source method \cite{allen1979image}. An illustration of the geometric setup is given in Figure \ref{fig:simudata}. All rooms have a rectangular shape with their dimensions and reverberation characteristics, described by the $\text{T}_{60}$ reverberation time, uniformly sampled from the ranges given in Table \ref{table:simuparams}. We use a circular microphone array with three omni-directional microphones. The diameter of the microphone array is $10$ cm. With respect to the x and y axis, we position the microphone array randomly in the room, however with a minimum distance of $1.2$ m to the walls. The height of the array is fixed at $1.5$ m. As illustrated in Figure \ref{fig:simudata}, the microphone array rotation is denoted by $\varphi_m \in [0, 2\pi)$.

During training, the spatially selective filter learns to extract a single target speaker from the mixture given its angle $\varphi_t$. The target speaker is represented by a red cross in Figure \ref{fig:simudata}. Its corresponding \ac{DoA} angle $\varphi_t$ is measured with respect to the microphone orientation as indicated by the blue dashed line. Interfering speakers are placed in the gray area at a minimum distance of $0.8$ m and a maximum distance of $1.2$ m just like the target speaker itself. We leave a  $10^\circ$ space with no interfering speakers around the target speaker as indicated by the white space in Figure \ref{fig:simudata}. The area of the gray annulus is divided into equally spaced segments as indicated by the gray dotted lines and one interfering speaker is randomly placed per segment, also with a minimum angular distance of $10^\circ$ to speakers in a neighboring segment. In Figure \ref{fig:simudata}, the interfering speakers are marked by a black cross. The height of the speakers is sampled from a normal distribution with mean $1.6$ m and standard deviation $0.08$ m.  

For training the spatially selective filter, we use a setup with five interfering speakers as shown in the Figure \ref{fig:simudata}. We discretize the target speaker location $\varphi_t$ using a $2^\circ$ resolution, which results in 180 target speaker directions and provide 300 examples per direction. This results in a total of $54,000$ training examples.

For training or testing on a speech separation task, we do not change any simulation parameter, but may vary the number of \say{interfering speakers}, which are then also considered as additional target speakers. For validation and testing, we generate $2,700$ and $1,800$ examples respectively.  The dry clean speech utterances are taken from the WSJ0 dataset \cite{wsjdata2007}, with no overlap between training, validation and test datasets. The sampling rate is $16$ kHz. The average \ac{DRR} for each individual speaker's signal is $-0.8$ dB and $95$ \% of the samples lie in the interval $[-5.9,~ 4.8]$ dB. For a separation scenario, we can characterize the distribution of the \ac{SNR} with respect to all included speaker extraction tasks considering one speaker as target signal and the mixture of interfering speakers as noise. In our setup, the \ac{SNR} is mainly influenced by the number of interfering speakers and the distance of the speakers to the microphone array. For two speakers, the \ac{SNR} of the extraction tasks range between $[-9.4,~9.4]$ dB for $95$\% of the data. For three and five speakers, the separation problem gets more difficult as the \ac{SNR} ranges shift to $[-11.8,~4.9]$ dB and $[-14.5,~0.5]$ dB respectively.

\section{Evaluation of the Separation Performance}\label{sec:sepperf}
\begin{table*}[]
    \caption{Speech separation performance for reverberant mixtures of two, three and five speakers. We compare an approach based on a spatially selective filter (SSF) with a direct separation (DS) approach using the same network architecture: JNF \cite{tesch2022interspeech, tesch2022tasl} or McNet \cite{yang2023mcfuse}.}
    \centering
    \begin{tabular}{l@{\hspace{0.2cm}}lcl@{\hspace{0.2cm}}ccccccccc}\toprule
    &&\multirow{2}{*}{DoA}& \multicolumn{3}{c}{2 speakers} &  \multicolumn{3}{c}{3 speakers} & \multicolumn{3}{c}{5 speakers}  \\ \cmidrule(lr){4-6}  \cmidrule(lr){7-9} \cmidrule(lr){10-12}
    No.&&& $\Delta$POLQA& $\Delta$SI-SDR & DNSMOS & $\Delta$POLQA& $\Delta$SI-SDR & DNSMOS & $\Delta$POLQA& $\Delta$SI-SDR & DNSMOS \\\midrule
    1 &JNF-DS & -- & 1.20 &11.7& 2.80& 0.87& 11.5 & 2.46 & 0.53& 10.7& 2.11\\\arrayrulecolor{black!30}\midrule %
    2 &JNF-SSF & oracle& 1.41& 12.7& 2.94& 1.30& 14.2& 2.79& 0.96& 15.1 & 2.52\\ %
    3 &JNF-SSF & search& 1.40 &  12.6&  2.94 & 1.29& 13.9 & 2.78 & 0.93& 14.4&2.51\\\arrayrulecolor{black!70}\midrule
    4 &McNet-DS & -- & 1.82 & 15.0& 3.03& 1.40& 15.4& 2.79& 0.87& 14.2& 2.39\\
    5 &McNet-iDS & oracle & 1.82 & 15.7& 3.07& 1.61& 15.9& 2.85& 0.96& 15.0& 2.43\\\arrayrulecolor{black!30}\midrule%
    6 & McNet-SSF & oracle& 1.85&14.7& 3.13& 1.76& 16.3& 3.04& 1.43& 17.3 & 2.84\\
    7 &McNet-SSF & search& 1.91& 15.0& 3.15& 1.80&16.3 & 3.06& 1.43& 16.6 & 2.85\\ %
    8 &McNet-SSF & DNN& 1.85& 14.7& 3.13& 1.76& 16.2& 3.04 & 1.42 & 16.9& 2.84\\\arrayrulecolor{black!70}\midrule
    9 & MVDR + PF & oracle & 0.42& 3.8 & 2.47& 0.23 & 2.8& 2.20& 0.14&3.1& 1.90\\
    10 & McNet-SSF (HCF) & oracle &1.49&11.6&2.90&1.38&12.6&2.78&1.03&12.4&2.53\\
    \arrayrulecolor{black}\bottomrule %
    \end{tabular}
    \label{tab:sepresults}
\end{table*}

A well-performing multi-channel speech separation system can be expected to benefit from knowledge of the speakers' locations. This assumption has led researchers to propose a variety of direction-based features, which are used as additional inputs to enhance \ac{DNN}-based speech separation \cite{wang2018mcdeepclustering ,gu2020spatialfeature, wang2019spatialspectralfeatures}. While a localization might happen implicitly in a regression-based direct separation (DS) approach, the spatially selective filter (SSF) separates the localization from the speaker extraction task and puts a strong focus on the spatial properties as the networks learns to focus on a single speaker using its direction as cue. In this section, we aim to investigate the impact of the chosen method, DS or SSF, on the separation performance. 

To ensure a fair comparison, we use the same underlying network structure, JNF and McNet as described in Section \ref{ssec:netarc}. For the DS approach, we omit the conditioning mechanism shown on the right side of Figure \ref{fig:jnfarc} and Figure \ref{fig:mcnetarc} and only provide the multi-channel mixture STFT as input. The output dimension of the last layer is changed so that not only one mask is predicted but as many masks as there are speakers in the mixture. We assume that the number of speakers is known. The network then produces an estimate for every speaker, and we use the same $\ell_1$ loss in time and frequency domain as for the SSF, but we apply it in a \ac{PIT} \cite{kolbaek2017pit} scheme. The loss function and other DNN training settings are described in Appendix \ref{app:trainingdetails}.

In Table \ref{tab:sepresults}, we report the separation performance for mixtures of two, three and five speaker mixtures measured using the \ac{POLQA} score \cite{polqa2018} improvement, \ac{SI-SDR} \cite{roux2019sisdr} improvement and DNSMOS \cite{reddy2022dnsmos} score. The \ac{POLQA} and DNSMOS score predict values on a \ac{MOS} score ranging from one (bad) to five (excellent). In contrast to \ac{POLQA} and \ac{SI-SDR}, DNSMOS does not require a reference signal, but is a neural network trained on user ratings according to the P.835 standard. We report the overall quality rating. 

Row number 1 displays the results for the DS networks based on the JNF architecture and trained with \ac{PIT}. Separate networks have been trained for the different numbers of speakers. In contrast, all results in row 2 have been obtained with the same network implementing the SSF approach based on the JNF architecture and evaluated given oracle \ac{DoA} information for the individual speakers. As the speakers are likely not positioned on the $2^\circ$ grid that has been used during training, we map the oracle target speaker direction onto the closest point in the grid before computing the one-hot encoding to condition the network as described in Section \ref{ssec:conditioning}. A comparison of the first two rows of Table \ref{tab:sepresults} reveals that the  JNF-SSF outperforms the JNF-DS approach in all metrics and for all numbers of speakers in the mixture. Furthermore, we observe that the performance difference is larger for a higher number of speakers in the mixture. For example, the POLQA performance difference between JNF-DS and JNF-SSF increases from 0.21 for two speakers to 0.43 for three and five speakers. 

Both networks JNF-DS and JNF-SSF used for the comparison in this table have approximately the same number of parameters. However, since the SSF is evaluated multiple times with each speaker as the target, the DS approach has a smaller number of learnable parameters per speaker. To investigate the influence of this effect, we also train JNF-DS with an increased number of parameters. For this we scale both LSTM layers by the same factor. The F-LSTM then has 364, 448 and 576 units for two, three and five speakers. The results are shown in Table \ref{table:jnf_ds_hidden_perf}. Comparing with row 1 in Table \ref{tab:sepresults}, we find that increasing the network size improves the performance of JNF-DS. However, as can be seen from the second block of Table \ref{table:jnf_ds_hidden_perf}, the JNF-SSF still outperforms JNF-DS. We therefore conclude that the superiority of the SSF approach can not be solely explained by an increased amount of parameters per speaker.

\begin{table}[tb]
    \caption{Speech separation performance of JNF-DS with the number of network parameters scaled according to the number of speakers to extract and the performance of JNF-SSF with multiple evaluations of the same network.}
    \label{table:jnf_ds_hidden_perf}
    \centering
\begin{tabular}{lccccc}\toprule
    &  \# Speakers & Param. [M] & $\Delta$POLQA& $\Delta$SI-SDR & DNSMOS \\\midrule\addlinespace[0.5em]
    \multirow{3}{*}{\rotatebox[origin=c]{90}{JNF-DS}}&2  & 2.4 & 0.83&10.5&2.82\\
    &3  & 3.6 & 1.07 & 13.2 & 2.62\\
    &5  & 6.0 & 0.70 & 12.7 & 2.28\\\midrule\addlinespace[0.5em]%
    \multirow{3}{*}{\rotatebox[origin=c]{90}{JNF-SSF}}&2  & 1.2$\times$2 & 1.41& 12.7&2.94\\
    &3  & 1.2$\times$3 & 1.30 & 14.2 & 2.79\\
    &5  & 1.2$\times$5 & 0.96 & 15.1 & 2.52\\[0.75ex]
    \bottomrule
\end{tabular}
\vspace{-1.5em}
\end{table}

\begin{figure}
    \centering
    \includegraphics{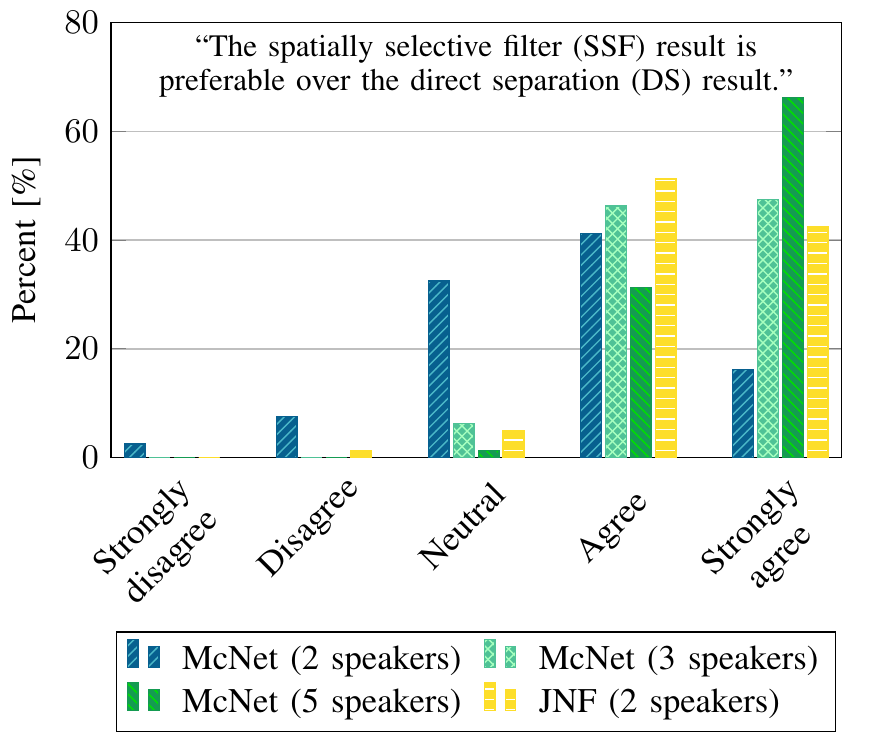}
    \caption{Results for a listening experiment assessing the participants' preference for separation results obtained with a spatial filter (SF) or a direct separation (DS) result. Speaker locations are assumed to be known for the spatial filter. The test is conducted blindly without test subjects knowing which example corresponds to which algorithm. The results have then been aggregated to match with the displayed statement.}
    \label{fig:listening_pref}
\end{figure}

As expected, the extension of the JNF architecture to McNet leads to a significant performance improvement in all metrics and for both the DS and SSF configuration. Comparing row 4 and 6, we find that the performance of McNet-SSF for two speaker mixtures is only slightly better than McNet-DS considering the POLQA score improvement and DNSMOS. It is even  worse by $0.3$ dB with respect to the SI-SDR measure. However, for more speakers the previously observed trend that the SSF output performs the DS network persists. For example, the performance difference amounts to 0.36 and 0.56 POLQA improvement score for three and five speakers, which is in line with a strong preference for the SSF result in a listening experiment as shown in Figure \ref{fig:listening_pref}. In this listening experiment, which evaluates the listener's preference for the SSF or DS result. Ten test subjects have been asked to rate a statement regarding their preference for one of two separation results obtained with the SSF or DS method. We use eight examples for each of the four comparisons. Of course, the test has been conducted blindly without a labeling of the methods and a random order of the compared items. The statement to be rated has the form: \say{Example 1 is preferable over Example 2.} We then aggregated the results to comply with the statement displayed in Figure \ref{fig:listening_pref}. While the metric scores in Table \ref{tab:sepresults} for two speaker mixtures are quite similar for McNet-DS and McNet-SFF, we observe a clear preference of the test subjects towards the SSF result. In total, more than $55$\% of the SSF test examples have been rated to be preferable over the corresponding DS result, which is favored only for $10$\% of the examples. Please find audio examples on our website\footnote{\url{https://uhh.de/inf-sp-ssf}}.

\begin{figure*}[htb]
    \centering
    \includegraphics{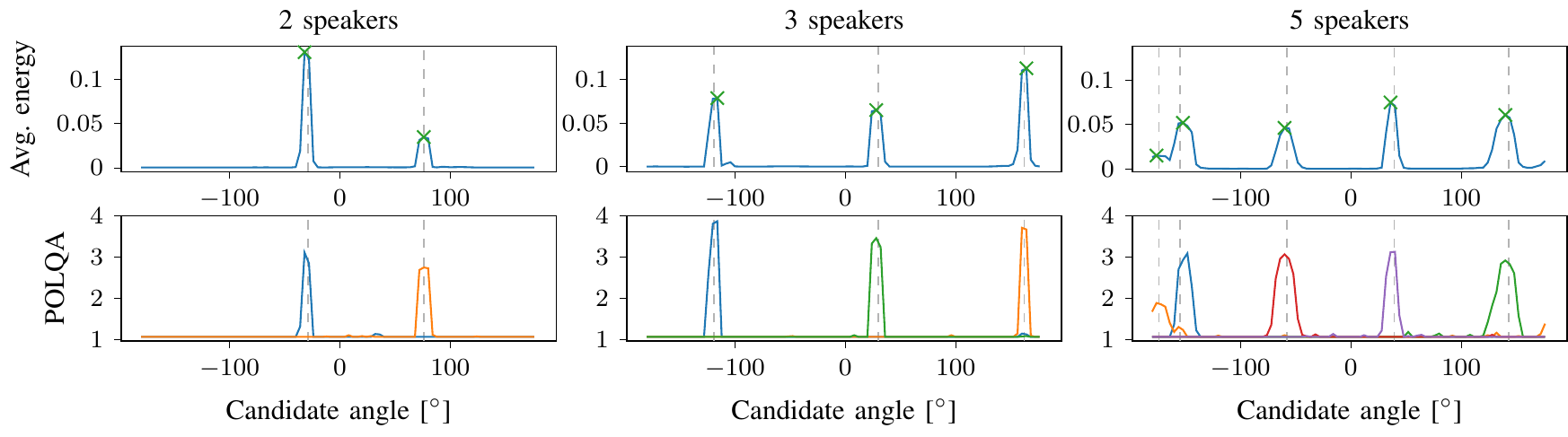}
    \caption{Examples for blind speaker separation and localization by peak-searching for a mixture of two, three and five speakers using non-linear filters steered in all candidate directions. The vertical dashed gray lines indicate the true positions of the speakers and the green cross marks the speaker location estimated based on the energy peaks in the filter output.}
    \label{fig:peaks}
    \vspace{-0.5em}
\end{figure*}

The SSF results discussed so far, have been obtained using oracle knowledge about the speaker \ac{DoA} angles. For classic beamforming, e.g. with an \ac{MVDR} beamformer, it is well-known that errors in the speaker \ac{DoA} estimates used to construct a steering vector are likely to cause significant performance degradation \cite{li2004robustmvdr}. To get an idea of the performance that can be expected in a blind setting, we also report results for two different \ac{DoA} estimation strategies in Table \ref{tab:sepresults}. The sensitivity to errors in the \ac{DoA} estimates is then investigated in more detail in Section \ref{sec:doaerrors}. The first strategy is search-based and evaluates the SSF for a set of potential target directions. The results for the individual speakers are then selected based on the energy of the filtered signal. The search-based strategy is illustrated in Figure \ref{fig:peaks}. The top row shows the energy of the filtered signals evaluated on a grid with $4^\circ$ resolution. We compute the energy on $10$ ms long non-overlapping segments and plot the average energy for all segments, in which speech is active. A $-45$ dB threshold with respect to the maximum energy in the mixture signal is used for detection of speech activity as in \cite{gerkmann2012noisepower}. We observe clear peaks at the true speaker locations, which are marked with a dashed gray line. The green crosses visualize the \ac{DoA} angle estimates, which have been obtained with a peak-finding heuristic applied to the energy curve of which the details are outlined in Appendix \ref{app:peaks}. 

Comparing the vertical dashed lines and the position of the green crosses in Figure \ref{fig:peaks}, we observe that the search-based \ac{DoA} estimates are quite accurate for the selected examples. The evaluation of the localization accuracy shown in Table \ref{table:doa_acc} asserts that these examples can be considered representative of the dataset, since also the mean angular error is low. As the search-based \ac{DoA} estimates depend on the DNN-based filter output, we display separate results for the JNF-SSF and the McNet-SSF in the first and second row. Somewhat unexpectedly, the mean angular error of the search-based \ac{DoA} estimates is smaller for JNF-SFF than McNet-SSF. However, in both cases the estimates are accurate enough to replace the oracle \ac{DoA} information without a loss in separation performance as can be seen in rows 3 or 7 of Table \ref{tab:sepresults}. For the McNet architecture the results obtained with the search-based strategy are even slightly better than those obtained with oracle \ac{DoA} information. While this observation might raise questions at the first sight, it can be explained by the second row of Figure \ref{fig:peaks}, which shows POLQA scores. We compute the POLQA measure for each candidate location's McNet-SSF output with respect to one speaker's reference signal to obtain one of the colored curves. The plots clearly show that the spatial filter has a high spatial selectivity for its steering direction, which is also why the search-based \ac{DoA} estimation works well. Consider now the first peak in the left-most plot. The \ac{DoA} estimate denoted by a green cross is slightly off to the left. However, the POLQA score at this estimated position is higher than the POLQA score at the true position. As a result, slight deviations in the search-based \ac{DoA} estimation are not harmful to the overall performance, but can even be helpful as they are correlated with the filter's behavior. As we will show in Section \ref{sec:doaerrors}, an uncorrelated \ac{DoA} error of $2^\circ$ causes a performance degradation. 

\newcolumntype{m}[1]{D{:}{\pm}{#1}}
\newcolumntype{C}{ @{}>{${}}c<{{}$}@{} }
\begin{table}[tb]
    \caption{The speaker localization accuracy for mixtures of two, three and five speakers in a reverberant room. We report the mean angular error in degree and the 95\% confidence interval.}
    \label{table:doa_acc}
    \label{tab:loc_acc}
    \centering
\begin{tabular}{l *3{rCl} }
    \toprule
     DoA estimation& 
     \multicolumn{3}{c}{2 speakers} & \multicolumn{3}{c}{3 speakers} & \multicolumn{3}{c}{5 speakers}\\\midrule
    search (JNF-SSF)  & 1.57   & \pm& 0.12                & 2.06  & \pm & 0.19    & 3.54 & \pm & 0.25\\
    search (McNet-SSF) & 2.07   & \pm& 0.07                & 2.53  & \pm & 0.15    & 3.99 & \pm & 0.23\\
    DNN & 1.06  & \pm & 0.03 & 1.24 & \pm & 0.09            & 2.13 & \pm & 0.19\\\bottomrule
\end{tabular}
\end{table}

Even though it provides interesting insights in the spatial selectivity of the proposed SSF, the search-based approach is too computationally demanding for most realistic applications. Therefore, we also evaluate the SSF with DoA estimates provided by a DNN-based classifier, which is trained to detect for every $2^\circ$ bin if there is a speaker or not. More details are provided in Appendix \ref{app:classifier}. Table \ref{table:doa_acc} shows that the DNN-based classifier is not only much more efficient, but also outperforms the search-based strategy in terms of localization accuracy by up to $1.86^\circ$ mean angular error for five speakers. The separation results for McNet-SSF are given in row number 8. Also for these DNN-based DoA estimates, we do not see major deviations from the oracle performance, which demonstrates that the SSF approach is well applicable also to blind separation tasks.

The two bottom rows provide results for two baseline systems. The first one is a traditional \ac{MVDR} beamformer with a DNN-based post-filter (PF). The parameters of the \ac{MVDR} are estimated from oracle data. The time-varying noise covariance matrices are estimated by recursive averaging of the pure noise data and the time-invariant \ac{RTF} estimate is obtained by multiplying the principal eigenvector of the generalized eigenvalue problem for speech and noise covariance matrices with the speech covariance matrix as described in \cite{ito2017}. The post-filter is a single-channel DNN with two LSTM layers trained on \ac{MVDR} outputs as described in \cite{tesch2022tasl}. The comparison with JNF-SSF and McNet-SSF highlights that drastic improvements can be achieved by replacing the linear spatial filter with a non-linear one. While the former does not achieve a substantial performance improvement over the mixture recording in a setting with three microphones and five speakers, the DNN-based SSFs provide respectable performance also in this difficult scenario. 

The second baseline uses hand-crafted features (HCF). For a fair comparison, we use the same McNet architecture as before but exchange the input signal. Previously, we provided the network with raw frequency domain data, i.e., the real and imaginary parts stacked. Following the approach proposed in \cite{gu2019neuralspatialfilter}, we replace this input by a compilation of features: the real and imaginary part of the reference channel as spectral feature, the \acp{IPD} between all pairs of microphones and the location-guided angle feature \cite{2018chenlocationguided, gu2019neuralspatialfilter}, which provides information about which speaker to extract. Comparing the results in row 10 with those in row 6, we find that our proposed approach without using hand-crafted features and with explicit conditioning on the one-hot encoded angle is beneficial. It outperforms the baseline with hand-crafted features by about 0.4 POLQA improvement score for two, three and five speakers. The performance benefit is particularly audible for three and five speaker mixtures. 

A question that might come to mind is whether the measured performance difference can be attributed to the difference in the approach (explicit spatial filtering in SSFs versus implicit spatial filtering in DS) or whether the explicit provisioning of \ac{DoA} information introduces a bias since speaker DoA information is generally known to be helpful for a multi-channel speaker separation problem. To investigate this, we train the DS networks again, but now also provide additional \ac{DoA} information for all speakers in the mixture. For this, we use the conditioning mechanism shown in Figure \ref{fig:netarc} using a multi-hot encoding.

We report the DoA-informed direct separation (iDS) results using the McNet architecture in the fifth row of Table \ref{tab:sepresults}. For two speakers, the performance obtained with the DoA-informed network is quite similar to the performance of McNet-DS in row 4 of Table \ref{tab:sepresults}. For three and five speakers, the additional \ac{DoA} information leads to a performance improvement over the McNet-DS.  However, the performance is still worse than that of McNet-SSF for three speakers and five speakers by 0.11 and 0.47 POLQA score respectively. Thus, providing the \ac{DoA} information does not close the performance gap between McNet-DS and McNet-SSF, especially not for a large number of speakers.

\section{Robustness Experiments}\label{sec:robustness}
For an approach to be applicable in practice, it is not only the maximum performance that is of interest, but also the robustness of the system, which is its capability to tolerate perturbations, for example in the geometric setup of the microphone array. In the following subsections, we evaluate the robustness of the SSF approach with respect to errors in the \ac{DoA} estimates and variances in the microphone array setup. All experiments in this and the next section have been conducted using the McNet architecture. Data generation parameters that are not explicitly mentioned in the experiment description are kept constant at the value or range of values specified previously in Section \ref{sec:data}.

\subsection{DoA Estimation Errors}\label{sec:doaerrors}
\begin{figure}
    \centering
    \includegraphics{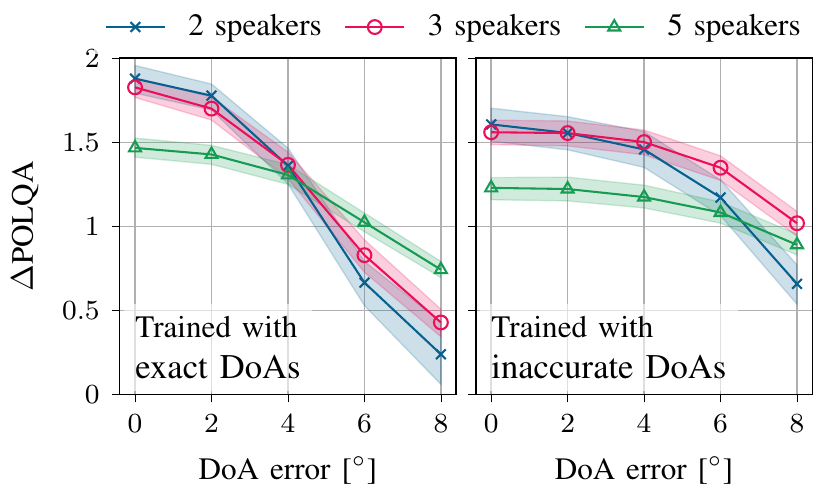}
    \caption{Separation results for the McNet-SSF conditioned on a target angle that is subject to a localization error of varying magnitude. During evaluation, the respective error is added to all speakers' DoA angles. The results shown in the left plot are obtained with with a McNet-SSF that has been conditioned on the exact DoA location during training, while the results displayed on the left side are obtained with a McNet-SSF that has been trained with inaccurate DoAs that include an error of up to $4^\circ$.}
    \label{fig:loc_err}
\end{figure}
Figure \ref{fig:loc_err} displays the results of an experiment that investigates the sensitivity of the SSF to errors in the \ac{DoA} estimates, which are used to steer the filter. For this, we add an error term to the oracle \ac{DoA} angle of all speakers in the mixture. The magnitude of the error is displayed on the x-axis. The y-axis represents the POLQA improvement score obtained with the McNet-SSF approach. We report average results and the $95\%$ confidence interval for 100 randomly selected mixture signals. The left plot shows the results for McNet-SSF, which has been trained using the exact DoA angle of the target speaker. As expected, the performance decreases as the DoA error, which is added only during evaluation, increases. On the positive side, though, the performance drop is not very drastic for a small error of $2^\circ$. In this case the performance loss is about $5\%$, $7\%$ and $3\%$ for two, three and five speaker mixtures respectively. Furthermore, we note that the more difficult problem with five speakers in the mixture is less affected by an \ac{DoA} estimation error. This observation can be explained by looking at the peak plots in Figure \ref{fig:peaks}. Here we can see in the bottom row that the peaks for five speakers have become wider and, therefore, an erroneous \ac{DoA} estimate to steer the filter has less consequences. 

The right side of Figure \ref{fig:loc_err} displays results for a McNet-SSF trained with inaccurate DoA information. During training we add a DoA error of up to $4^\circ$ to the true DoA. As a consequence, we observe that the performance stays approximately the same for a DoA error of up to $4^\circ$ degrees. The performance drop for larger errors is also significantly reduced for two and three speakers. However, this increased robustness comes at the cost of a reduced performance if the DoA estimates are accurate, which is expected to some extend as there will always be a trade-off  between sensitivity to \ac{DoA} estimation errors and the spatial selectivity of the filter. In line with this, we observe that the peaks reflecting the spatial selectivity as in Figure \ref{fig:peaks} have widened for the filter trained with inaccurate DoAs.

\subsection{Perturbations in the Microphone Placement}
\begin{figure}
    \centering
    \includegraphics{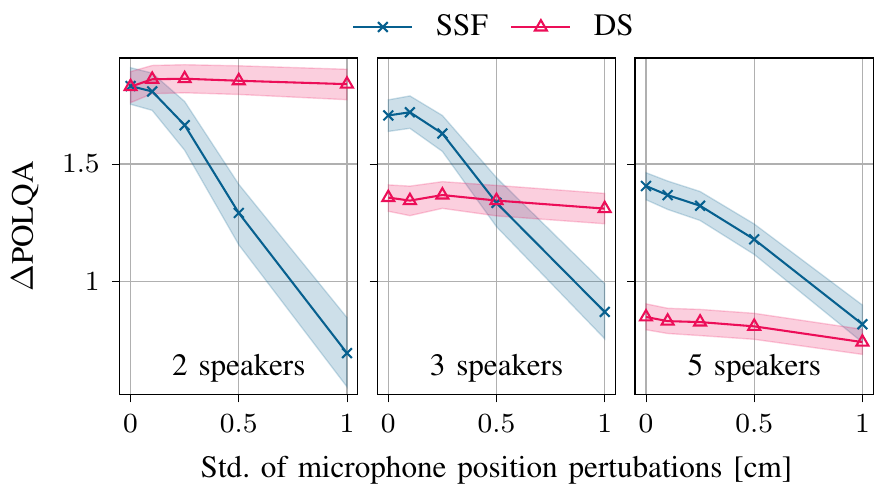}
    \caption{Separation results for test examples with pertubated microphone positioning. The noise added to the microphone placement is sampled from a zero-mean normal distribution with a standard deviation shown on the x-axis.}
    \label{fig:mic_pert}
\end{figure}

In a multi-channel scenario, the spatial information is very closely related to the geometric configuration of the microphone array, since the main source of information is the relative \acp{TdoA} of a signal at the microphones. So far, we have used a fixed and exact placement of the microphones in the array for generation of the simulated training and testing data. Now we add some noise to the positioning of the microphones to evaluate the sensitivity of the SSF and DS networks to these kinds of perturbations. We sample the noise that we add to the x-, y- and z-coordinate of the microphones in the circular array from a zero-mean normal distribution with a standard deviation between $0.1$ cm and $1$ cm. The standard deviation, i.e. the amount of perturbation in the microphone array geometry, is shown on the x-axis of Figure \ref{fig:mic_pert} against the POLQA improvement score. 

The plots from left to right in Figure \ref{fig:mic_pert} show the results for two, three and five speaker mixtures. Clearly, the SSF approach is highly sensitive to perturbations in the microphone array geometry. While small perturbations with a standard deviation of $1$ mm are tolerated without a significant loss in performance, large perturbations render the method useless. In contrast, the DS performance only slightly decreases for five speakers and is approximately constant in the other cases. From a perspective of robustness this can be considered a favorable behavior. However, the DS performance for the three and five speaker mixtures is far from the maximum performance reached by the SSF method even with some perturbations in the microphone array geometry. Therefore, we would not consider the DS approach robust to variations in the spatial characteristics related to the microphone array geometry but view the results of this experiment as a strong indication that the DS approach performs worse than the SSF approach for a higher number of speakers because it does not fully exploit the spatial information in the multi-channel data in the first place.

\section{Generalization Experiments}\label{sec:generalization}
As neural networks learn to extract patterns from data, for example the general spectro-temporal structure of speech or the spatial characteristics of a signal arriving from a particular direction, it is insightful to evaluate the performance of a DNN-based approach also for inputs whose characteristics vary from those present in the training set. Here we perform an experiment that varies the distance between speech source and microphone array, an experiment investigating the performance for speech sources with similar \ac{DoA} and one that adds an additional noise source.

\subsection{Far-field vs Near-field Scenario}

\begin{figure}
    \centering
    \includegraphics{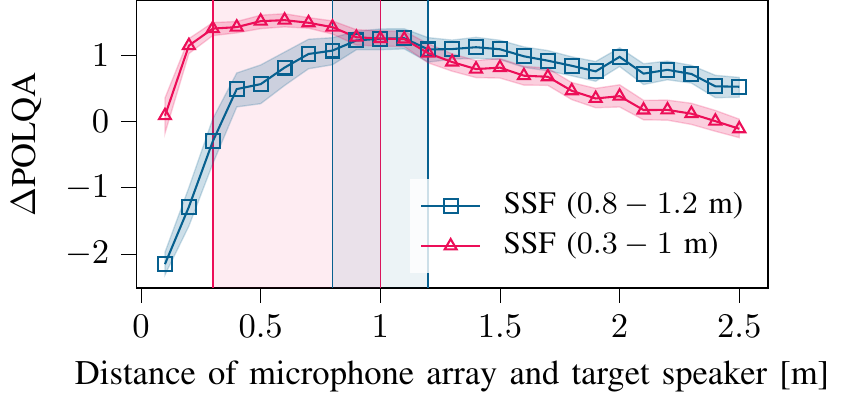}
    \caption{POLQA improvement scores for a single speaker placed at a varying distance to the microphone array. Results for MCNet-SSF trained with speakers at a distance of \mbox{$0.8-1.2$ m} are shown by the blue plot with square markers and the results of a MCNet-SSF trained with a target speaker positioned at a distance of $0.3-1.0$ m (as in \cite{tesch2023icassp}) are represented by red triangles. The range covered by the training data has been marked by the respective color. The diameter of the microphone array is $0.1$ m in both cases.}
    \label{fig:nf_vs_ff}
\end{figure}
The relative \acp{TdoA} of the direct-path signal are not only related to the microphone configuration in the array, but also influenced by the distance of the source to the microphone array. If the distance between the source and the microphone array is large compared to the distances between the microphones, we commonly make a far-field assumption and model the propagation of sound as a plane wave as opposed to spherical wave in the near-field \cite{vary2006digital}. We have trained the SSF with mixtures of sources that were placed at a distance of \mbox{$0.8-1.2$ m} to the microphone array, which itself has a diameter of \mbox{$10$ cm}. Here we investigate the performance of the SSF for input signals that deviate from the data seen during training with respect to their distance. For this experiment, we present the SSF network with a single reverberant speech signal originating from a varying distance and report the POLQA improvement score in Figure \ref{fig:nf_vs_ff}. The blue plot with square markers represents a network trained with data generated according to the configuration given in Section \ref{sec:data}. The range that is covered by the training data is shown be the blue shaded rectangle. We can see that the network reaches the maximum performance for examples that fall in this range. The performance gradually decreases as the source moves further away, which will also decrease the \ac{DRR}. If the source moves closer toward to microphone array on the left side of the blue area, the \ac{DRR} increases. Nevertheless, the performance drops even to negative improvement scores as the source moves very close to the microphone array. For our previous work on speaker extraction \cite{tesch2022interspeech, tesch2022tasl}, we have trained the SSF with a target source that is placed closer to the microphone array with a distance of $0.3-1.0$ m. The results for this filter are shown in red. Here we can see that including near-field examples in the training data drastically improves the performance for close-by sources. However, as the network spends more parameters modeling spatial characteristics in the near-field, the performance for far-distant sources is decreased.

\subsection{Sources with Similar \ac{DoA}}

\begin{figure}
    \centering
    \includegraphics{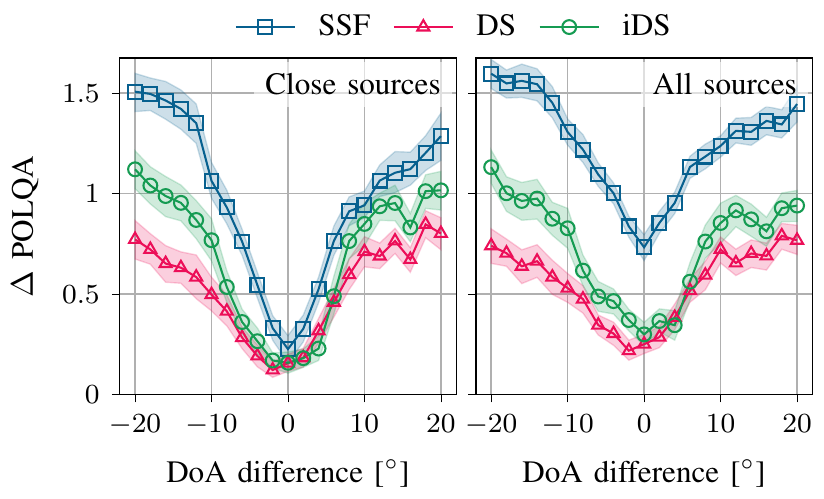}
    \caption{Separation results for test examples that include three speakers, of which one is placed at a $60^\circ$ angle, one at a $0^\circ$ angle and one speaker with at a variable angle between $-20^\circ$ and $20^\circ$. The left plot shows the average POLQA improvement for the two close sources and the right plot shows the average improvement for all three sources.}
    \label{fig:close_sources}
\end{figure}

All speech sources in the examples generated according to the dataset description in Section \ref{sec:data} have a minimum $10^\circ$ angle difference between them and every other source. 
While such such a constraint may be realistic for example in a meeting scenario with the microphone array positioned on the table, in other scenarios the speakers may stand closer together. We therefore investigate the generalization ability of the DS and SSF approach for close sources. For this, we generate test examples with mixtures of three speakers which are positioned at a $60^\circ$ angle, a $0^\circ$ angle and a variable angle between $-20^\circ$ and $20^\circ$. We evaluate on 60 examples for every angle difference. The average POLQA improvement for the two close sources is shown in the left plot of Figure \ref{fig:close_sources}. Clearly, neither the SSF nor the DS approach can provide good separation results for sources arriving from the same direction as can be seen by the performance dip at $0^\circ$ angle difference. However, listening to the results it becomes clear that they handle the task in different ways. For sources positioned in the same direction, the SSF returns a mixture of only the two close sources excluding the third speaker for both speakers, while the DS approach returns different results for every speaker, which are of very low quality.

The plot on the right side of Figure \ref{fig:close_sources} includes the separation result for the third speaker positioned at a $60^\circ$ angle when computing the average POLQA improvement score. Here we can see that for the SSF, shown in blue, the average POLQA improvement score increases by about $0.5$. In contrast, the average performance of the DS (red curve) remains about the same as in the left plot. This means that while the SSF is struggling to separate the two close sources, it has no problem to accurately extract the third source. On the other hand, the DS approach is not able to provide a reasonable result for the third speaker if two speakers are close. From these findings, we conclude that the decoupling of the separation results for the individual speakers in the SSF leads to a better generalization in a scenario that contains sources with similar \ac{DoA}. The green curve shows the performance for iDS, for which we provide the \ac{DoA} information as an additional input. Here we can see that such a desirable decoupling is not achievable by simply providing \ac{DoA} information to a network that is trained with \ac{PIT}.

\subsection{Noise (Unseen During Training)}
\begin{table}[]
    \caption{Separation results for two speaker mixtures with an additional music noise source at a random position. The number of sources refers to the number of sources predicted by each system. We then select the outputs corresponding to the two speech sources for evaluation.}
    \centering
    \setlength\tabcolsep{4.5pt}
    \begin{tabular}{llcccc}\toprule
    &DoA&\#Sources & $\Delta$POLQA& $\Delta$SI-SDR & DNSMOS  \\\midrule
    McNet-SSF& oracle & 2 & 1.66 & 15.2 & 2.98\\\arrayrulecolor{black!30}\midrule
    McNet-DS& & 2&  0.65& 6.5 & 2.28\\
    McNet-iDS& oracle& 2& 0.84& 10.0& 2.35\\
    McNet-DS& & 3& 1.29& 14.0 & 2.70\\
    McNet-iDS& oracle & 3 & 1.51 & 14.8& 2.80 \\\arrayrulecolor{black}\bottomrule
    \end{tabular}
    \label{tab:music_noise}
\end{table}

It is a major advantage of the SSF in comparison with the DS approach that the neural network does not need to be re-trained for evaluations on mixtures of different numbers of speakers as it focuses on extraction of speech signal from a particular direction. In our last experiment, we now add a music noise source to a mixture of two speakers and compare the ability of the SSF and DS approach to generalize to this new scenario. For the music noise source, we sample a random position between $0.8$ m and $1.2$ m away from the microphone array (same distance as speakers) and add the music signal to the mixture signal with an \ac{SNR} of $5$ dB. The music signals are taken from the \emph{jamendo} subset of the MUSAN dataset \cite{snyder2015musan}. 

The performance results are shown in Table \ref{tab:music_noise}. The first row shows the performance of the SSF given oracle knowledge of the two speakers' directions. Compared to the results in the fifth row of Table \ref{tab:sepresults}, the POLQA score improvement and DNSMOS score are reduced by about $10$\% and $5$\% respectively. However, the system still reliably separates the two speakers without the music signal leaking into one of the estimates. In contrast, McNet-DS trained on two speaker mixtures cannot handle the additional noise source, which is reflected by the low performance scores in the second row of Table \ref{tab:music_noise}. Results obtained with the \ac{DoA}-informed McNet-iDs are given in the third row. Interestingly, the performance notably improves if oracle knowledge about the speaker \acp{DoA} is provided, which was not the case for the noise-free scenario. However, the performance is still 0.82 POLQA and 5.2 dB SI-SDR lower than that of McNet-SSF. 

Since the music noise source could be seen as a third source, we also test the DS networks trained on three sources. For computing the performance results, we then only consider the outputs that correspond to the two speech sources. We can see in the two bottom rows of Table \ref{tab:music_noise} that this indeed improves the performance by a large margin. The best performance is obtained with McNet-iDS trained on three speech sources. However, there remains a performance gap of 0.15 POLQA score, 0.4 dB SI-SDR and 0.18 DNSMOS to the McNet-SSF, which, unlike McNet-iDS in the last row, does not require information about the noise source's \ac{DoA}. Since detecting the number of noise point sources as well as their \ac{DoA} will be difficult in most real-world scenarios, the SSF, which only needs \ac{DoA} estimates for the speech sources, is not only performing better but is also much more practical compared to all tested DS approaches.

\section{Conclusion}
Based on our conference paper \cite{tesch2023icassp}, we proposed a steerable DNN-based spatially selective filter (SSF). Beyond \cite{tesch2023icassp}, here we have investigated the separation performance of a DNN-based spatially selective filter (SSF) steered in the direction of each speaker in the mixture in comparison with a classic end-to-end direct separation (DS) approach trained with \ac{PIT}. We find that the SSF, which has been trained for high spatial selectivity in the given \ac{DoA}, outperforms a DS approach by a large margin if there are more than two speakers in the mixture. Experiments on the robustness of either system provides evidence that this is because the SSF better exploits spatial information. Furthermore, we have shown that the SSF generalizes much better to unseen noise conditions than the DS approach.

\appendix

\subsection{Network Training Details}
\label{app:trainingdetails}
All neural networks (JNF-DS, JNF-SSF, McNet-DS, McNet-iDS and McNet-SSF) have been trained on an $\ell_1$ loss in time and frequency domain \cite{tolooshams2020cadunet}:
\begin{equation}\label{eq:loss}
    L(s_p, \hat{s}_p) = \alpha\norm{s_p-\hat{s}_p}_1 + \norm{|S_p|-|\hat{S}_p|}_1.
\end{equation}
The frequency-domain term $\hat{S}_p$ is estimated as given in (\ref{eq:maskappl}) by multiplication of the noisy signal's reference channel with the network-estimated mask and $\hat{s}$ is its inverse \ac{STFT}. The parameter $\alpha$ is set to $\alpha=10$ to approximately equalize the contribution of either domain. We use the Adam optimizer \cite{KingmaB2015Adam} with an initial learning rate of 0.001 and reduce the learning rate by a factor $\gamma=0.75$ every $50$ epochs. We train for a maximum of 500 epochs and select the best weights based on the validation loss. We use $32$ ms windows for computing the \ac{STFT} with a $50$\% overlap and use the square-root Hann window for both analysis and synthesis. 

\subsection{Peak-finding Heuristic}
\label{app:peaks}
As a first step, we normalize the highest peak in the energy curve to one and then run \texttt{scipy.signal.find\_peaks} with a prominence of 0.009, a height of 0.05 and a width of one. If fewer peaks than the expected number of speakers are detected, we re-execute the function with relaxed parameters settings (no width requirement, decreased prominence of 0.001 and height of 0.025) and merge peaks that likely to represent the same speaker as they are close together and have a similar height. If more peaks than expected speakers are found, we pick the highest ones.

\subsection{DNN-based DoA Classifier}\label{app:classifier}
We train a DNN-based classifier for \ac{DoA} estimation, which is composed of an F-LSTM layer as described in Section \ref{ssec:netarc} and two feed-forward layers with 256 and 180 hidden units. We use an exponential linear unit activation for the first feed-forward layer and a sigmoid activation for the second. The network is trained to detect for every $2^\circ$ bin if there is a speaker or not based on the full utterance. We train for 100 epochs with an average binary cross-entropy loss on the dataset of two speaker mixtures. Even though trained on two speaker mixtures, the classifier performs sufficiently well in detecting other numbers of speakers. We use the same peak-finding heuristic as for the search-based strategy, which is necessary as the classifier provides an output between zero and one for every angle bin.

\section*{Acknowledgment}
We would like to thank J. Berger and Rohde\&Schwarz SwissQual AG for their support with POLQA.

\bibliographystyle{IEEEtran}
\bibliography{bib}

\begin{IEEEbiography}[{\includegraphics[width=1in,height=1.25in,clip,keepaspectratio]{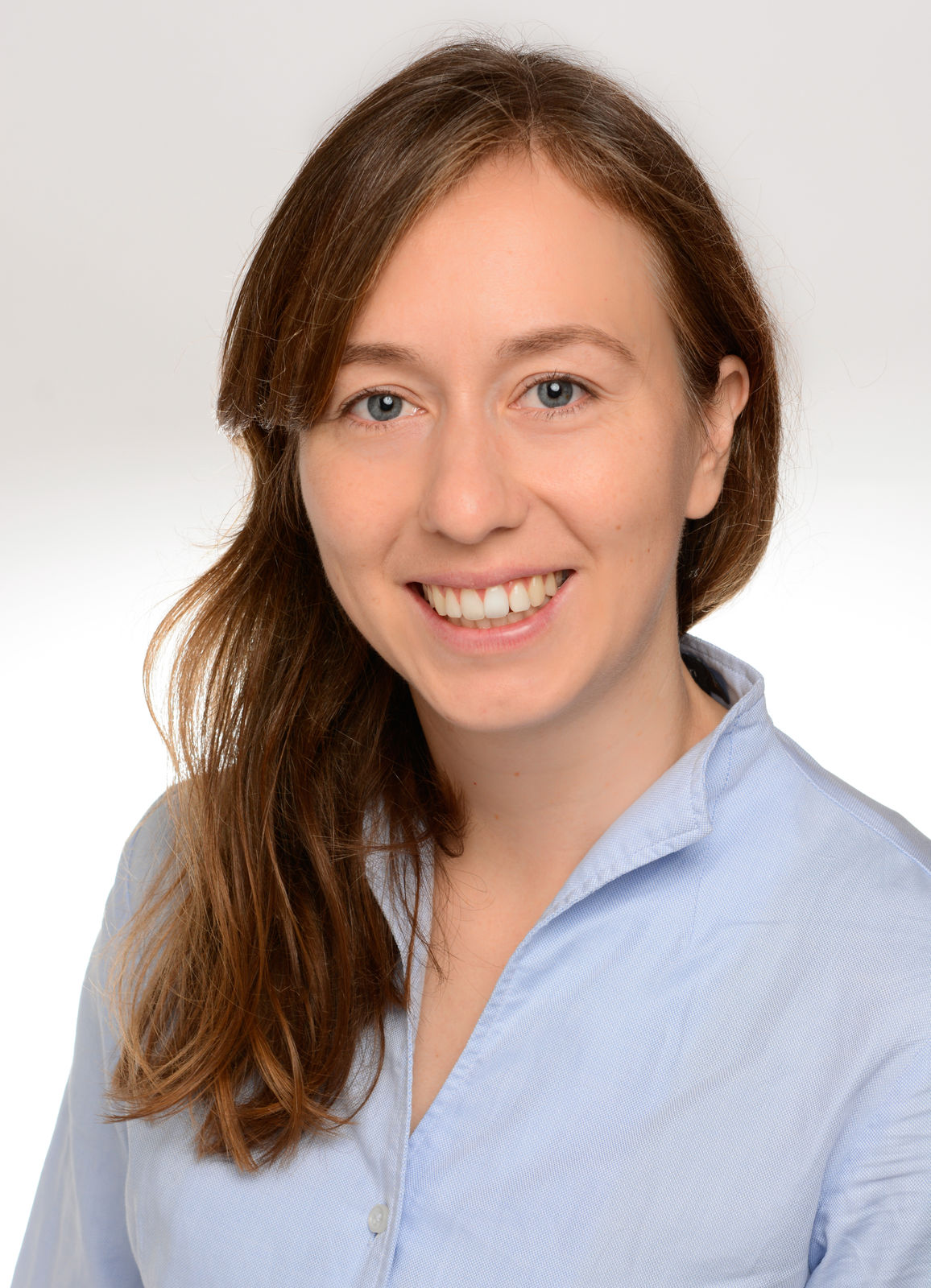}}]{Kristina Tesch}
(Student Member, IEEE) received a B.Sc and M.Sc in Informatics in 2016 and 2019 from the Universität Hamburg, Hamburg, Germany. She is with the Signal Processing Group at the Universität Hamburg since 2019 and is currently working towards a doctoral degree. Her research interests include digital signal processing and machine learning algorithms for speech and audio with a focus on multichannel speech enhancement. She received the ITG VDE award 2022 for the publication "Nonlinear Spatial Filtering in Multichannel Speech Enhancement" \cite{tesch2021nonlinearspatialfilteringtasl}. 
\end{IEEEbiography}

\begin{IEEEbiography}[{\includegraphics[width=1in,height=1.25in,clip,keepaspectratio]{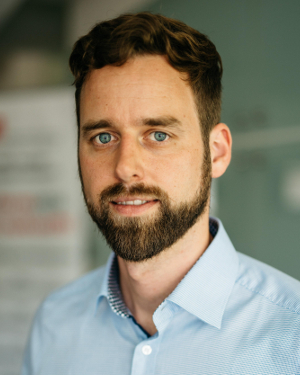}}]{Timo Gerkmann}
(Senior Member)
is currently a professor with the Universität Hamburg, Hamburg, Germany, where he is the head of the Signal Processing Research Group, and a deputy head of the Department of Informatics. He has previously held positions with Technicolor Research \& Innovation in Hanover, Germany, the Carl-von-Ossietzky Universität Oldenburg, Oldenburg, Germany, KTH Royal Institute of Technology, Stockholm, Sweden, Ruhr-Universität Bochum, Bochum, Germany, and Siemens Corporate Research in Princeton, NJ, USA. His main research interests include statistical signal processing and machine learning for speech and audio applied to communication devices, hearing instruments, audio-visual media, and human-machine interfaces. Timo Gerkmann is currently a Member of the IEEE Signal Processing Society Technical Committee on Audio and Acoustic Signal Processing and a Senior Area Editor of the IEEE/ACM Transactions on Audio, Speech, and Language Processing.
\end{IEEEbiography}

\end{document}